\documentclass[12pt]{article}
\usepackage{epsfig,a4,amssymb,amsmath}
\newcommand{\Dlr}{\stackrel{\leftrightarrow}{D}}

\newcommand{\pslash}{\not{\hspace{-0.08cm}p}}

\newcommand{\be}{\begin{equation}}
\newcommand{\ee}{\end{equation}}

\newcommand{\bea}{\begin{eqnarray}}
\newcommand{\eea}{\end{eqnarray}}
\newcommand{\MS}{{\overline{MS}}}

\title{
\vspace{-2.5cm}
\flushleft{\normalsize DESY 05-174} \\
\vspace{-0.35cm}
{\normalsize Edinburgh 2005/07} \\
\vspace{-0.35cm}
{\normalsize Leipzig LU-ITP 2005/019} \\
\vspace{-0.35cm}
{\normalsize Liverpool LTH 667} \\
\vspace{0.35cm}
\centering{\Large \bf Perturbative renormalisation of quark bilinear operators for overlap fermions with and without stout links and improved gauge action}
\footnote{Talk given by H. Perlt at the XXIIIrd International Symposium on Lattice Field Theory, 25-30 July 2005, Trinity College, Dublin, Ireland}\vspace*{-0.35cm}}
\author{\large R.~Horsley$^1$, H.~Perlt$^{2,3}$, P.E.L.~Rakow$^4$,
G.~Schierholz$^{5,6}$ and A.~Schiller$^3$ \\[0.75em]
-- QCDSF Collaboration -- \\[0.75em]
\normalsize
$^1$ School of Physics, University of Edinburgh, \\
\normalsize 
Edinburgh EH9 3JZ, UK \\
\normalsize
$^2$ Institut f\"ur Theoretische Physik, Universit\"at Regensburg,\\ 
\normalsize
D-93040 Regensburg, Germany \\
\normalsize
$^3$ Institut f\"ur Theoretische Physik, Universit\"at Leipzig, \\ 
\normalsize
D-04109 Leipzig, Germany \\
\normalsize
$^4$ Theoretical Physics Division, Department of Mathematical Sciences,\\ 
\normalsize
University of Liverpool,\\
\normalsize
Liverpool L69 3BX, UK \\
\normalsize
$^5$ John von Neumann-Institut f\"ur Computing NIC,\\
\normalsize
Deutsches Elektronen-Synchrotron DESY,\\
\normalsize
D-15738 Zeuthen, Germany\\
\normalsize
$^6$ Deutsches Elektronen-Synchrotron DESY, \\
\normalsize
D-22603 Hamburg, Germany
}

\date{ }

\begin{document}

\maketitle

\abstract{We calculate  lattice renormalisation constants of local and one-link quark operators
for overlap fermions and improved gauge actions in one-loop perturbation
theory.
For the local operators we stout smear the SU(3) links in the fermionic action.
Using the popular tadpole improved L\"uscher-Weisz actions at $\beta=8.45$ and
$\beta=8.0$ we present numerical values for the Z factors in the $\overline{MS}$
scheme (partly as function of the stout smearing strength).
We compare various levels of mean field (tadpole) improvement
which have been applied to our results.}


\section{Introduction}

Lattice calculations at small quark masses require an action with good chiral
properties. The same is true for calculations of matrix elements of certain
operators, which otherwise mix with operators of opposite chirality.
Ginsparg-Wilson fermions~\cite{Ginsparg:1981bj} have an exact chiral symmetry
on the lattice~\cite{Luscher:1998pq}, and thus are well suited for these
tasks. A further advantage is that they are automatically $O(a)$
improved~\cite{Capitani:1999uz}. Overlap fermions~\cite{overlap,n2,n} provide
a four-dimensional realisation of Ginsparg-Wilson fermions.

It is well known
that computer simulations with overlap fermions suffer from large
computational costs due to the large condition number. This number was
found to be reduced by using improved gauge actions~\cite{DeGrand:2002vu,Galletly:2003vf}
(even after projecting out the lowest eigenvalues; see however~\cite{Jansen:2003jq}). In
this paper we use the tadpole-improved  Luscher-Weisz
action~\cite{Luscher:1984xn,Luscher:1985zq,Alford:1995hw}
\begin{eqnarray}
S_G^{TILW} & =&\frac{6}{g^2} \,\,\Bigg[c_0
\sum_{\rm plaquette} \frac{1}{3}\, {\rm Re\, Tr\,}(1-U_{\rm plaquette})
\, + \, c_1 \sum_{\rm rectangle} \frac{1}{3}\,{\rm Re \, Tr\,}(1- U_{\rm
rectangle}) \nonumber \\
& &  + \, c_3 \sum_{\rm parallelogram} \frac{1}{3}\,{\rm Re \,Tr\,}(1-
U_{\rm parallelogram})\Bigg]\, .
\label{SGTILW}
\end{eqnarray}
The parameters $c_1$ and $c_3$ weight the contributions of the
corresponding six-links loops. It is customary to impose the
normalisation condition
\begin{equation}
c_0 + 8 c_1 + 8 c_3 = 1.
\label{normcond}
\end{equation}
Defining the lattice coupling $\beta$  by
\begin{equation}
\beta = \frac{6}{g^2} c_0
\end{equation}
we choose
in accordance with numerical simulations performed by the QCDSF
collaboration~\cite{Gurtler:2005,Streuer:2005}
the following values~\cite{Gattringer:2001jf}
\begin{equation}
\begin{tabular}{l|l|l}
$\beta$  & $c_1$ & $c_3$ \\
  \hline
&&  \\[-2ex]
$8.45$           & $-0.154846$  & $-0.0134070$\\ [0.7ex]
$8.0\phantom{0}$ & $-0.169805$  & $-0.0163414$
\end{tabular}
\end{equation}

A further step to reduce the computational costs consists of smearing
the gauge link variables in the fermionic action. The resulting fat links tend to reduce the
density of eigenmodes speeding up the inversion of the
fermion Dirac operator~\cite{DeGrand:2004nq,Durr:2005an}. Morningstar and
Peardon~\cite{Morningstar:2003gk} have proposed an analytic
smearing method (stout link method) which can be applied in
Hybrid Monte Carlo (HMC) simulations. As by construction the
stout links remain in the SU(3) group, this enables the force
term in the equations of motion for HMC to be easily
determined.

The calculations which we will present in the next sections have been
performed for $r=1$ and the overlap parameter $\rho=1.4$
(for the definition see~\cite{Horsley:2004mx}). The influence
of stout links is investigated for local operators only.

\section{Renormalisation}
\label{renormsection}

To obtain continuum results from lattice calculations of hadron matrix
elements, the underlying operators have to be renormalised. A non-perturbative
determination of the corresponding renormalisation factors would be
preferable. However, often perturbative renormalisations are  done first.
Especially, if this calculation is performed analytically it provides useful
information about the intrinsic singular structure and possible
complicated mixing properties. We use a {\it Mathematica} program
which has been developed for one-loop lattice perturbative
calculations~\cite{Gockeler:1996hg} and has been extended to
overlap fermions with improved gauge actions.

We define renormalised operators $\mathcal{O}$ by
\begin{equation}
\mathcal{O}^{\mathcal{S}}(\mu) = Z_{\mathcal{O}}^{\mathcal{S}}(a,\mu)
\mathcal{O}(a) \, ,
\end{equation}
where $\mathcal{S}$ denotes the renormalisation scheme. $Z_{\mathcal{O}}^{\mathcal{S}}(a,\mu)$
is the renormalisation factor connecting the lattice operator $\mathcal{O}(a)$
with the renormalised operator $\mathcal{O}^{\mathcal{S}}(\mu)$ at scale $\mu$.
We use the MOM scheme by computing the amputated Green function $\Lambda_{\mathcal{O}}$ of
the operator $\mathcal{O}$ and define the $Z_{\mathcal{O}}$ via
\begin{equation}
\label{ZMOM1}
\frac{Z^{MOM}_{\mathcal{O}}(a,\mu) }{Z^{MOM}_\psi(a,\mu)}\;
\Lambda_{\mathcal{O}}\; \big|_{p^2=\mu^2}
= \Lambda_{\mathcal{O}}^{\rm tree}
+{\rm \ other\  Dirac\  structures} \, ,
\end{equation}
where $Z^{MOM}_\psi(a,\mu)$ is the quark wave function renormalisation factor.
The renormalisation constants can be converted to the $\overline{MS}$ scheme,
\begin{equation}
\label{ZMOM2}
Z_{\psi,\mathcal{O}}^{\overline{MS}}(a,\mu) = Z_{\psi,\mathcal{O}}^{\overline{MS},MOM}
Z_{\psi,\mathcal{O}}^{MOM}(a,\mu)\,,
\end{equation}
where the conversion factors $Z_{\psi,\mathcal{O}}^{\overline{MS},MOM}$ are calculable
in continuum perturbation theory.

\section{One-link operators}
\label{onelinksection}

Let us first consider the one-link operators
\begin{eqnarray}
  {\cal O}_{\mu\nu} &=& \frac{i}{2}\,\bar{\psi} (x) \gamma_\mu \Dlr_\nu \psi(x) - {\rm traces}
\label{Omunu}\,,
  \\
  {\cal O}^5_{\mu\nu} &=& \frac{i}{2}\,\bar{\psi} (x) \gamma_\mu\gamma_5 \Dlr_\nu \psi(x) - {\rm traces}\,,
\end{eqnarray}
which are related to the first moments of
unpolarised and polarised nucleon structure functions,
where $\Dlr_\nu$ is the left-right covariant lattice derivative. The chiral properties of
overlap fermions imply that matrix elements of ${\cal O}_{\mu\nu}$ and ${\cal O}^5_{\mu\nu}$ give
identical results. Therefore, we restrict our calculations to the unpolarised case, i.e. to
${\cal O}_{\mu\nu}$. For a detailed discussion see~\cite{Horsley:2005jk}.

The amputated one-loop Green function $\Lambda_{\mu\nu}$ obtained from (\ref{Omunu}) has the form
\begin{eqnarray}
\label{fullstructure}
\Lambda_{\mu\nu} (a,p) &=&
\gamma_\mu p_\nu +
\frac{g^2 C_F}{16\pi^2}\bigg\{
 \left[ \left(\frac{1}{3} +\xi  \right) \log (a^2 p^2)  - 4.29201 \, \xi +b_1 \right]
\gamma_\mu p_\nu
\nonumber
\\
&&\hspace{-6mm}
+\left[ \frac{4}{3}  \log (a^2 p^2) + \frac{1}{2} \,  \xi +b_2 \right]
\gamma_\nu p_\mu
+\left[- \frac{2}{3}  \log (a^2 p^2) - \frac{1}{2} \,  \xi +b_3 \right]
\delta_{\mu\nu} \pslash
\nonumber
\\
&&\hspace{-6mm}
+\,b_4 \, \delta_{\mu\nu} \gamma_\nu p_\nu  +\left( -\frac{4}{3}+\xi\right) \frac{p_\mu p_\nu}{p^2} \pslash
\bigg\}\,,
\end{eqnarray}
with $\xi$ as gauge parameter (Feynman gauge: $\xi=0$) and $C_F=4/3$. The constants $b_i$ are
linear combinations of finite lattice integrals~\cite{Gockeler:1996hg} and depend on
the used lattice fermionic and gauge actions. For the chosen actions and parameters we obtain
\begin{equation}
\begin{tabular}{c|c|c|c|c|c}
Action  & $b_1$ & $b_2$& $b_3$& $b_4$ & $b_\Sigma$\\
  \hline
&&&&&  \\[-2ex]
$\beta=8.45$ & $-5.6115$  & $-3.8336$ &  $2.7793$ & $0.3446$  & $-16.180$\\ [0.7ex]
$\beta=8.0$  & $-5.2883$  & $-3.7636$ &  $2.7310$ & $0.3331$  & $-15.733$
\end{tabular}
\label{structtab}
\end{equation}
In (\ref{structtab}) we have added the contribution of the quark self energy $b_\Sigma$
which is needed for the calculation of the wave function renormalisation.

{}From (\ref{ZMOM1}), (\ref{ZMOM2}) and (\ref{fullstructure}) we determine the Z factors in the $\MS$ scheme for the commonly used
representations under the hypercubic group
\begin{eqnarray}
\tau_3^{(6)}: \quad {\cal O}_{v_{2a}}  \equiv  \frac{1}{2}
\left({\cal O}_{14}+ {\cal O}_{41}  \right) \,,
\quad\quad
\tau_1^{(3)}: \quad {\cal O}_{v_{2b}}  \equiv  {\cal O}_{44} -\frac{1}{3}
\left({\cal O}_{11}+ {\cal O}_{22}+{\cal O}_{33}  \right) \nonumber\,.
\end{eqnarray}
They have the form
\begin{equation}
Z_{v_{i}}^{\MS}(a,\mu)  =
  1 - \frac{g^2 C_F}{16 \pi^2}
  \left[\frac{16}{3}\log(a\mu)+ B_{v_{i}}(c_k,\rho)  \right]
\end{equation}
with
\begin{equation}
B_{v_{i}}(c_k,\rho) =\frac{40}{9} + b_{v_{i}} + b_\Sigma\,,
\quad b_{v_{2a}} = b_1+b_2\,,\quad b_{v_{2b}} = b_1+b_2+b_4 \,.
\end{equation}

It is well known that the naive perturbative results suffer from lattice
artefacts. Therefore, mean field (tadpole) improvement~\cite{Lepage:1992xa} has been
proposed to rearrange the perturbative series. In case of overlap
fermions the tadpole improved Z factor is given by~\cite{Horsley:2004mx}
\begin{equation}
Z_\mathcal{O}^{TI} = Z_\mathcal{O}^{MF}
\left(\frac{Z_\mathcal{O}}{Z_\mathcal{O}^{MF}}\right)_{\rm pert}\,,
\end{equation}
where $Z_\mathcal{O}^{MF}$ is the mean field approximation of
$Z_\mathcal{O}$. For overlap fermions we have
\begin{equation}
Z_{\cal O}^{MF} =  \frac{\rho}{\rho -4 (1- u_0)},\quad
Z_{{\cal O}{\rm pert}}^{MF} =  1 + \frac{g^2_{TI}\,C_F}{16\pi^2}\frac{4}{\rho}k_u^{TI}\,.
\end{equation}
Here $u_0$ denotes the mean value of the link. The boosted parameters are chosen as~\cite{Horsley:2004mx}
\begin{equation}
g_{TI}^2 =g^2/u_0^4 \,,\quad
c_0^{TI} = c_0 \,, \quad
c_i^{TI} = u_0^2 \,c_i \,(i=1,3), \quad
C_0^{TI} = c_0 + 8 c_1^{TI} + 8 c_3^{TI}\,.
\end{equation}
$k_u^{TI}$ is the one-loop contribution of the perturbative expansion for $u_0$
with $c_i^{TI}$ inserted for the corresponding gauge actions
($k_u^{TI} = 5.3625/5.0835$ for $\beta=8.45/8.0$).

In case of overlap fermions one needs to improve the quark propagator as well which leads
to a boosted $\rho$ parameter
\begin{equation}
\rho^{TI}=\frac{\rho - 4(1-u_0)}{u_0} \,.
\end{equation}
Results obtained with $\rho^{TI}$ are denoted as fully tadpole improved (FTI).
The Z factors have the form
\begin{equation}
Z_{v_{i}}^{TI,\MS} =Z_{\cal O}^{MF}
 \left\{1 - \frac{g_{TI}^2 C_F}{16 \pi^2} \left[
  \frac{16}{3\,C^{TI}_0} \log(a\mu)
  + B_{v_{i}}^{TI}(c_k^{TI},\rho) \right]\right\}\,,
\end{equation}
\begin{equation}
Z_{v_{i}}^{FTI,\MS} = Z_{\cal O}^{MF}
 \left\{1 - \frac{g_{TI}^2 C_F}{16 \pi^2} \left[
  \frac{16}{3\,C^{TI}_0} \log(a\mu)
  + B_{v_{i}}^{TI}(c_k^{TI},\rho^{TI}) \right]\right\}\,.
\end{equation}
The following table shows numerical results for the various levels of
improvement at $a=1/\mu$
\begin{equation}
\label{tab3}
\begin{tabular}{c|c|c|c|c|c|r|c}
Operator  & $\beta$ & $B$ & $Z^\MS$& $B^{TI}$& $Z^{TI,\MS}$ & $B^{FTI}$& $Z^{FTI,\MS}$\\
  \hline
  &&&&&&&\\[-2ex]
  $v_{2a}$    &  $8.45$   & $-22.430$  &$1.315$ & $0.502$
                &$1.393$        &$-0.077$ & $1.411$\\[0.7ex]
  $v_{2b}$    &  $8.45$   & $-22.085$  &${\bf 1.311}$ & $0.793$
                &${\bf 1.384}$       &$0.230$ & ${\bf 1.401}$\\[0.7ex]
		\hline
&&&&&&&\\[-2ex]
  $v_{2a}$    &  8.0 & $-22.036$  &$1.310$        & $0.603$
                &$1.390$       &$-0.108$ & $1.412$ \\[0.7ex]
  $v_{2b}$    &  8.0 & $-21.703$  &$1.305$       & $0.892$
                &$1.381$       & $0.199$ & $1.402$
\end{tabular}
\end{equation}
It can be read off from Table (\ref{tab3}) that the one-loop corrections $B$
for the improved perturbative Z factors become smaller as expected.
Thus the perturbative series is better behaved. For $\beta=8.45$ and representation
$v_{2b}$ we can compare the perturbative Zs (bold faced numbers in (\ref{tab3})) with a quenched Monte Carlo simulation~\cite{Gurtler:2005p} giving $Z^{MC,\MS}=1.98(3)$. Using the stout smearing procedure (see Section \ref{localsection})
the resulting factors are $Z^{MC,\MS}_{1-smear}=1.47(4)$ and
$Z^{MC,\MS}_{2-smear}=1.34(3)$ with the value of the smearing parameter
$\omega=0.15$.

\section{Local operators and stout smearing}
\label{localsection}

Z factors for local fermionic operators for overlap fermions and a set of
improved gauge actions have been determined in~\cite{Horsley:2004mx}. A recalculation
has been performed by Ioannou and Panagopoulos~\cite{Panagopoulos:2005}.
In this section we show the influence of stout smearing on the perturbative
Z factors.

By construction a stout smearing step is performed on a gauge link variable $U_\mu(x)$
as~\cite{Morningstar:2003gk}
\begin{equation}
\label{stout1}
	U_\mu^{(n+1)}(x) = e^{iQ_\mu^{(n)}(U,\omega_{\mu\nu})}\,U_\mu^{(n)}(x),
\end{equation}
where $n$ denotes the step of smearing. The parameters $\omega_{\mu\nu}$ characterise the
strength of smearing: they are the weights of the perpendicular staples
associated to the link $(x,x+\hat{\mu})$.
For our perturbative calculation we have assumed
the isotropic case
$\omega_{\mu\nu}=\omega$ and a small value of
$\omega$. Various investigations suggest  values of $0.1 < \omega <
0.3$. Therefore, we have expanded (\ref{stout1}) to first order in $\omega$.
Furthermore, we have restricted ourselves to $n=1$. The resulting stout link has been inserted
into the fermionic action modifying the corresponding Feynman rules for
the quark-gluon vertices. The corresponding results are obtained in powers
of $\omega$. As a possible choice in the tadpole improvement we assumed
that our approximate stout smearing has been done for the mean field rescaled
links (this does not change $Z_{\cal O}^{MF}$ and does not rescale $\omega$).

As examples we have calculated the Z factors in  $\MS$-scheme for the scalar and axial vector
operators for the TILW action at $\beta=8.45$ and the perturbative improvement
levels discussed in the previous section. For the scalar operator we find ($a=1/\mu$)
\begin{eqnarray}
Z_S&=&1.168 - 0.248 \,\omega - 0.154 \,\omega^2\,,\nonumber\\
Z_S^{TI}&=&1.309 - 0.488 \,\omega - 0.239 \,\omega^2\,,\nonumber\\
Z_S^{FTI}&=&1.359 - 0.241 \,\omega - 0.685 \,\omega^2\,.\nonumber
\end{eqnarray}
These numbers can be compared with $Z^{MC}$ obtained from a quenched MC
simulation at $\omega=0.15$ and a single smearing~\cite{Gurtler:2005p}:
$$
Z_S = 1.127\,,  Z_S^{TI} = 1.230\,,  Z_S^{FTI} = 1.307\,;
 Z_{S,0-smear}^{MC}=1.36(1)\,,  Z_{S,1-smear}^{MC}=1.13\,.
$$
The same has been done for the axial vector operator. We get in this case
\begin{eqnarray}
Z_A&=&1.156 - 0.475 \,\omega + 0.092 \,\omega^2\,,\nonumber\\
Z_A^{TI}&=&1.268 - 0.860 \,\omega + 0.179 \,\omega^2\,,\nonumber\\
Z_A^{FTI}&=&1.303 - 0.560 \,\omega - 0.346 \,\omega^2\,.\nonumber
\end{eqnarray}
The comparison with MC gives for $\omega=0.15$
$$
Z_A = 1.087\,, Z_A^{TI} = 1.114\,, Z_A^{FTI} = 1.211\,;
 Z_{A,0-smear}^{MC}=1.42(1)\,, Z_{A,1-smear}^{MC}=1.16\,.
$$
Contrary to the non-perturbative case the perturbative stout smearing
decreases only mildly the renormalisation factors.

\section*{Acknowledgements}

The authors thank Stephan D\"urr for useful comments concerning
new results for speeding up the overlap calculations.
This work is supported by DFG under contract FOR 465 (Forschergruppe
Gitter-Hadronen-Ph\"anomenologie).

\end{document}